# Ultrafast laser-induced spin-lattice dynamics in the van der Waals antiferromagnet CoPS$_3$

D. Khusyainov[1], T. Gareev[1], V. Radovskaia[1], K. Sampathkumar[1,2], S. Acharya[3], M. Šiškins[4], S. Mañas-Valero[6], B.A. Ivanov[1,5], E. Coronado[6], Th. Rasing[1], A.V. Kimel[1] and D. Afanasiev[1]

[1]Radboud University, Institute for Molecules and Materials, 6525 AJ Nijmegen, The Netherlands.

[2] Central European Institute of Technology, Brno University of Technology, Purkyňova 648/125, Brno, 62100 Czech Republic

[3]National Renewable Energy Laboratory, Golden, CO 80401, USA

[4]Kavli Institute of Nanoscience, Delft University of Technology, Lorentzweg 1, 2628 CJ, Delft, The Netherlands

[5]Institute of Magnetism, NAS and MES of Ukraine, 36b Vernadsky Blvd., Kiev 03142, Ukraine

[6]Instituto de Ciencia Molecular (ICMol), Universitat de València, c/Catedrático José Beltrán 2, 46980, Paterna, Spain

**Abstract.**

*CoPS$_3$ stands out in the family of the van der Waals antiferromagnets XPS$_3$ (X=Mn, Ni, Fe, Co) due to the unquenched orbital momentum of the magnetic Co$^{2+}$ ions which is known to facilitate the coupling of spins to both electromagnetic waves and lattice vibrations. Here, using a time-resolved magneto-optical pump-probe technique we experimentally study the ultrafast laser-induced dynamics of mutually correlated spins and lattice. It is shown that a femtosecond laser pulse acts as an ultrafast heater and thus results in the melting of the antiferromagnetic order. At the same time, the resonant pumping of the $^4T_{1g} \rightarrow {}^4T_{2g}$ electronic transition in Co$^{2+}$ ions effectively changes their orbital momentum, giving rise to a mechanical force that moves the ions in the direction parallel to the orientation of their spins, thus generating a coherent $B_g$ phonon mode at the frequency of about 4.7 THz.*

## Introduction

Since the seminal discovery of ultrafast demagnetization in Ni[1], the ultrafast manipulation of magnetism with ultrashort pulses of light has evolved into a fascinating research topic of nonequilibrium magnetism, with examples ranging from excitation of collective magnetic modes[2–5] to light-driven magnetic phase transitions[6–8] and switching of spin orientation[9–11]. The recent resurgence of interest in two-dimensional (2D) van der Waals (vdW) materials hosting intrinsic long-range magnetic orders has offered a novel playground for investigating these phenomena in systems where the interplay between structural and magnetic orders plays a pivotal role[12–16]. Understanding the nonequilibrium dynamics in vdW magnets particularly promises to provide important insights into the fundamental physics of spin-lattice interactions and spin relaxation in ultimately thin magnets. Moreover, the inherently strong light-matter interactions, typical of vdW materials[12,17–23], open up exciting possibilities for efficient manipulation of magnetism on the ultrafast timescale, promising novel energy-efficient data processing devices for future spintronics and magnonics applications[24,25].

Among the various vdW magnets, transition metal thiophosphates, $XPS_3$ ($X$ = Fe, Ni, Mn, and Co), form a unique class of 2D antiferromagnets (AFMs) with intralayer AFM order on a honeycomb lattice. In addition to the 2D AFM order, previous studies of $XPS_3$ have uncovered strongly coupled spin and charge orders[26,27], highly anisotropic excitons[28], magneto-electric coupling[29,30], signatures of the BKT transition[31] and strong electron correlations[32,33]. The family has also recently attracted a lot of attention for ultrafast control of magnetism. In particular, it has been shown that a sudden perturbation of electron orbital momentum via resonant pumping of specific electronic transitions opens up new ways to control spins and lattice at an ultrafast timescale[18,19,33,34]. $CoPS_3$ stands out in the $XPS_3$ family due to the $Co^{2+}$ ions, characterized by the large spin and unquenched orbital momentum. The presence of such ions in magnetic materials usually results in a strong coupling of the spins to the lattice[35,36], large magnetocrystalline anisotropy[37], high frequencies of magnetic resonance[38], and strong photomagnetic effects[10,35,36]. Furthermore, it has (not sure about "been")recently suggested that the high-spin $d^7$ configuration of $Co^{2+}$-based compounds may host a dominant Kitaev interaction[39–41]. However, despite intense studies of the $XPS_3$ family, ultrafast optical control of magnetism and lattice in $CoPS_3$ remains unexplored.

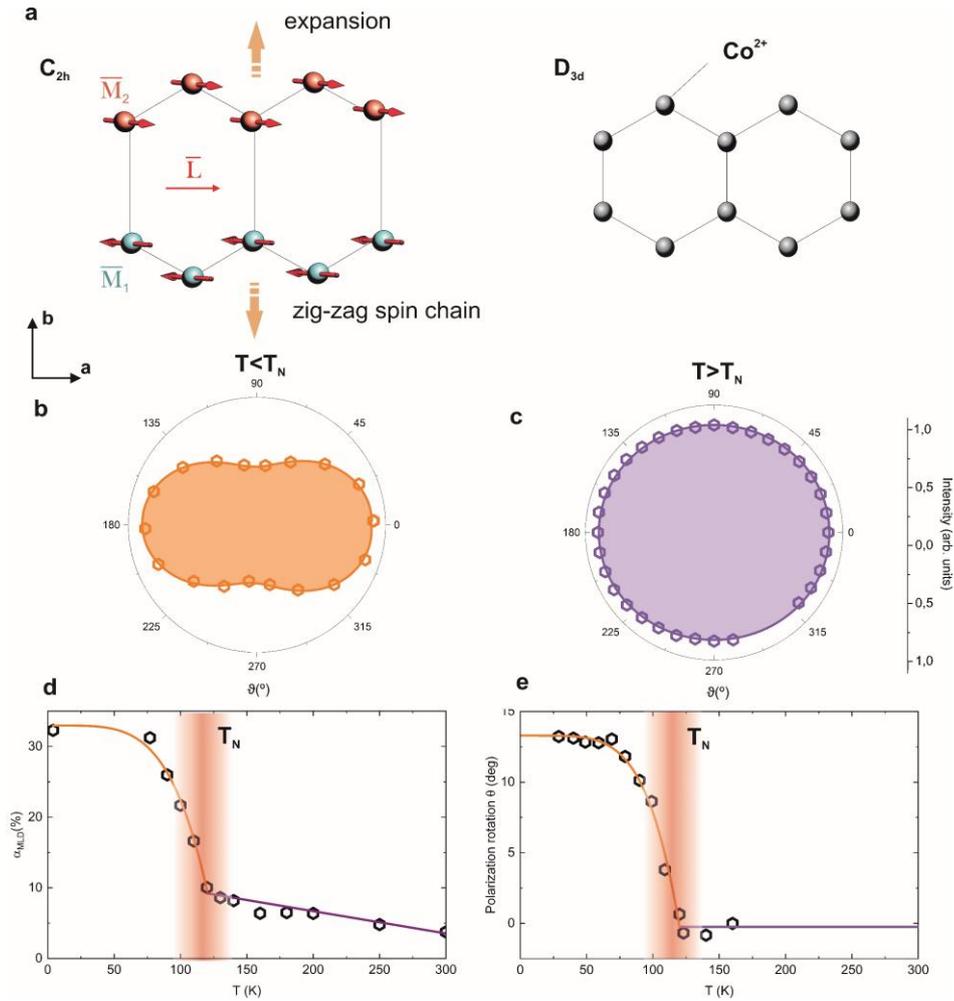

**Figure 1.** *(a) Schematics of the AFM spin and crystal structure of $CoPS_3$ in the antiferromagnetic and paramagnetic phases. (b,c) The normalized transmission of light as function of linear polarization angle $\vartheta$ measured below and above $T_N$, respectively. The angle $\vartheta=0$ corresponds to the orientation of the polarization plane along the a-axis. (d) Magnetic linear dichroism ($\alpha_{MLD}$) of $CoPS_3$ measured as a function of temperature. Solid line is a guide to the eye. (e) Rotation of the light polarization plane $\theta$ as a function of temperature. Solid line is a guide to the eye. The photon energy of the probe light is 1.55 eV.*

Here we address these shortcomings by first introducing optical magnetic linear dichroism (MLD) as an efficient means to probe the AFM order in $CoPS_3$ and then employing time-resolved magneto-optical pump-probe spectroscopy to detect the ultrafast light-induced dynamics of spins and lattice in this AFM compound. Comparison of the results with earlier studies for MnPS$_3$ and FePS$_3$ reveals that the presence of the large unquenched orbital momentum in CoPS$_3$ results in a substantially higher MLD and much faster laser-induced melting of the spin order. Moreover, the unquenched momentum leads to exceptionally strong spin-lattice coupling and provides an opportunity for highly efficient optical control of the lattice dynamics by selective pumping of specific orbital transitions in magnetic Co$^{2+}$ ions.

## Sample and Experimental Procedure

CoPS$_3$ is a layered van der Waals AFM, characterized by a weak coupling between the adjacent crystal and magnetic layers, that can be viewed approximately as a quasi-2D antiferromagnet[42]. The intralayer spin ordering below $T_N$=120 K results in the formation of ferromagnetic "zig-zag" chains along the *a*-axis, while the adjacent chains in the direction of the *b*-axis are AFM coupled (see Fig 1a). Introducing individual magnetizations of the adjacent chains as **M**$_1$ and **M**$_2$, the AFM order in CoPS$_3$ can be naturally characterized by a Néel vector **L**, such that **L** = **M**$_2$-**M**$_1$. Inelastic neutron scattering shows that CoPS$_3$ has sizeable easy-axis single-ion anisotropy which defines the orientation of the spins and the resultant Néel vector along the crystallographic *a*-axis[39]. The formation of spin chains, combined with a strong spin-lattice interaction of Co$^{2+}$ ions, leads to a reduction of the crystal symmetry, such that the point group of a single CoPS$_3$ layer changes from $D_{3d}$ to $C_{2h}$ [15]. Above $T_N$ in the paramagnetic (PM) phase, the crystal lattice is characterized by a six-fold rotational symmetry. Below $T_N$, the ordering of spins in chains leads to a compression of *a* and elongation of *b* lattice parameters, effectively elongating the hexagons formed by the Co$^{2+}$ ions in the direction perpendicular to the spin chains. The structural changes that accompany the magnetic transition strongly affect the vibrational (phonon) spectrum of CoPS$_3$. Recent Raman studies show that upon spin ordering several otherwise double degenerate E$_g$ phonon modes lose their degeneracy and split into a pair of nondegenerate A$_g$ and B$_g$ ones, the frequencies of which demonstrate anomalous behavior below $T_N$[15].

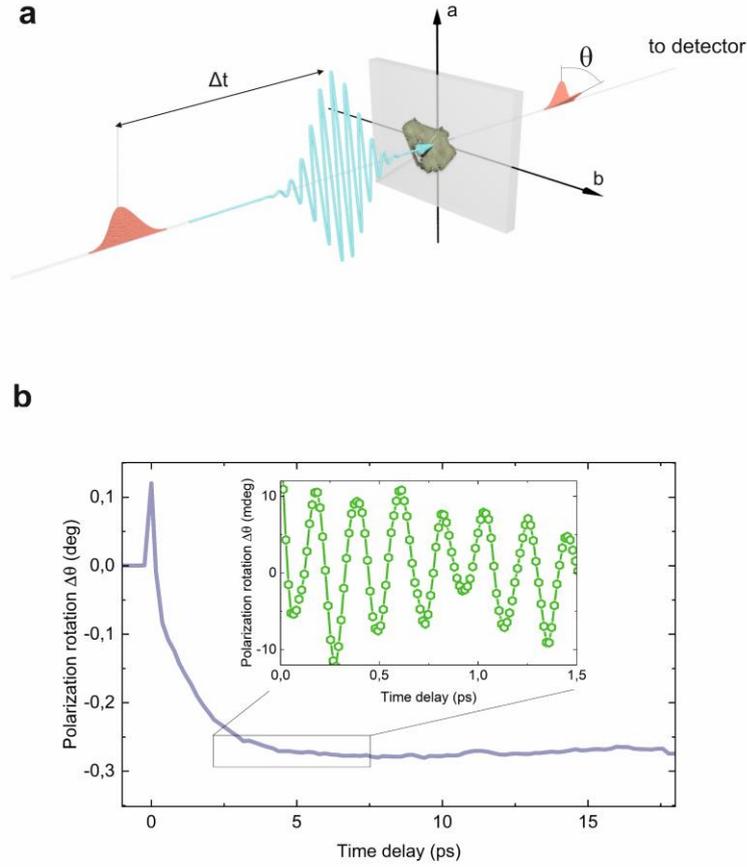

**Figure 2. (a)** *Schematics of the time-resolved magneto-optical pump-probe experiment in $CoPS_3$. The pump pulse (blue) is delayed with respect to the probe pulse by a variable time delay Δt.* **(b)** *Time-resolved rotation of the probe pulse polarization plane exemplifying both pump-induced quenching of the AFM order and excitation of the coherent $B_g$ phonon mode (inset). The sample temperature is set to 77 K.*

To be able to employ an all-optical pump-and-probe technique for studies of ultrafast magnetism in $CoPS_3$, one must find an efficient method of optical detection of spin order in the AFM phase. Earlier it was reported that several $XPS_3$ materials possess a large magnetic linear dichroism (MLD) in the AFM phase[12,43]. Moreover, according to Ref.[12], the strength of the dichroism seems to scale with the orbital moment of the magnetic ion. The MLD is the smallest in $MnPS_3$, where the orbital momentum of the magnetic $Mn^{2+}$ ions is quenched, and the largest in $FePS_3$, where the orbital momentum of $Fe^{2+}$ is known to be non-zero. To reveal the strength of this magneto-optical effect in $CoPS_3$, with an even larger orbital momentum associated with the $Co^{2+}$ ions, we measured the transmittance of linearly polarized probe light at the photon energy of 1.55 eV as a function of the angle $\vartheta$ between the electric field of light and the *a*-axis of the crystal. Figures 1b, and 1c show the polarization dependencies below ($T=7$ K) and above ($T=300$ K) the $T_N$. It is seen that the normalized transmittance is strongly anisotropic if the material is in the AFM state. Measuring the difference between the intensities $I_a$ and $I_b$ of the transmitted light polarized along the *a*- and the *b*-axes, respectively, we define MLD as:

$$\alpha_{\mathrm{MLD}} = \frac{(I_a - I_b)}{(I_a + I_b)} \cdot 100\%. \tag{1}$$

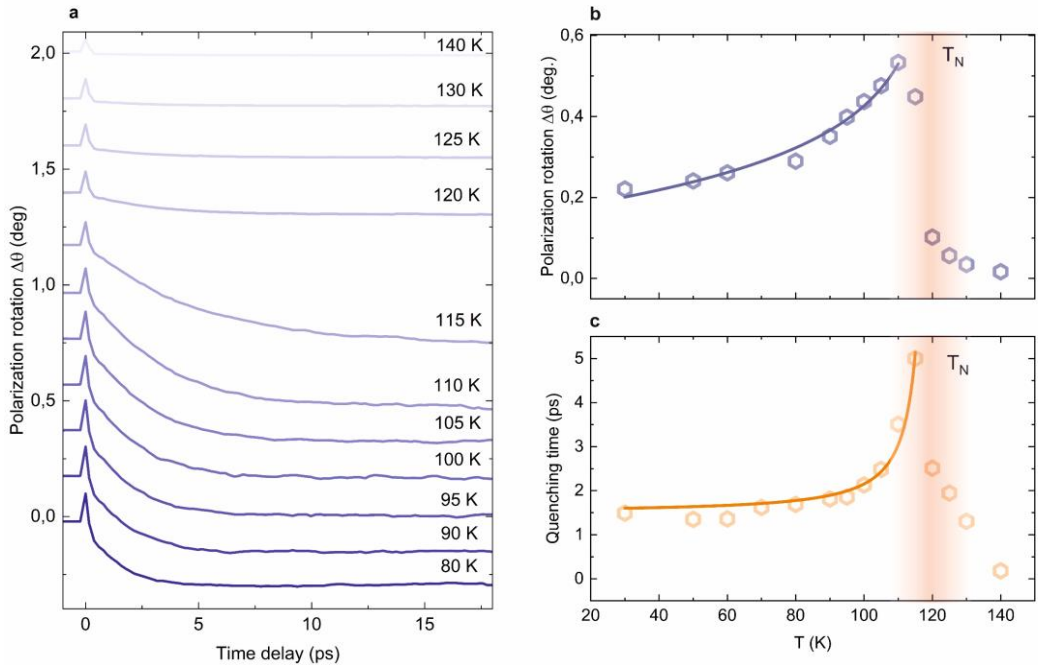

**Figure 3 | Ultrafast light-induced quenching of the Néel order in CoPS$_3$** *(a) Time-resolved rotation of the probe polarization Δθ showing pump-induced quenching of the AFM order for various temperatures across $T_N$. The pump photon energy is 0.9 eV. (b) The magnitude of the quenching as a function of temperature T. (c) Quenching time as a function of T.*

Although MLD, in general, strongly depends on the probe photon energy, already in our experiment its value reaches 30%. This is almost an order of magnitude larger than the largest MLD reported for FePS$_3$[12]. To define the origin of the MLD in CoPS$_3$ we studied the effect as a function of the probe photon energy. As shown in Supplementary Materials Fig. S1, two well-pronounced MLD bands appear with the maxima centered at the photon energies of about 1.69 and 2.34 eV. These energies closely match the energies of the so-called *d-d* transitions of Co$^{2+}$ ions that are responsible for the change of the orbital state of the magnetic ion[44]. This observation thus indicates that the *d-d* transitions are the origin of MLD in CoPS$_3$, in agreement with previous studies[12,18,19,34]

Figure 1d shows MLD in CoPS$_3$ as a function of temperature *T*. It is seen that $α_{MLD}$ reduces significantly when approaching $T_N$ from below. To quantify the critical behavior of $α_{MLD}$, we fitted the temperature dependence of $α_{MLD}$ in the AFM phase using a power law $α_{MLD} \propto |T_N - T|^{2β}$. The results, shown in Fig. 1d, yield $2β = 0.60 \pm 0.01$. This value agrees well with the critical exponent characterizing a temperature scaling of the decay in the intensity of the neutron Bragg diffraction peak from the underlying AFM order in CoPS$_3$ which was found to be $2β = 0.60$ [42]. As the Bragg intensity scales with the Néel order as $L^2$, its comparison with our results suggests that $α_{MLD}$ also scales as $L^2$. In theory, MLD should be quadratic concerning **L**, but experimentally it is not always the case, especially in AFMs with strong piezomagnetism[45]. At the same time, MLD is known to be one of the most universal effects to probe the AFM order using light in a broad spectral range, including THz[46], visible[47,48], and X-rays[49]. Yielding $β ≈ 0.30$ for the critical exponent characterizing the temperature dependence of the AFM order parameter, our results further confirm that the spin order in CoPS$_3$ is best described by the 3D Ising model[50]. Note, the dichroism, as defined in this work, does not depend on the sample thickness, and thus can be potentially employed to probe AFM spin order in CoPS$_3$ down to a single layer. At the same time, we observed that in the finite size CoPS$_3$ samples studied in our experiments, the linear dichroism does not completely vanish above $T_N$, see Fig. 1d. This is because the ideal six-fold rotational symmetry of the honeycomb lattice, present in a single-layer form, vanishes in bulk crystals due to the displacement of the stacked layers along the *a*-axis[30].

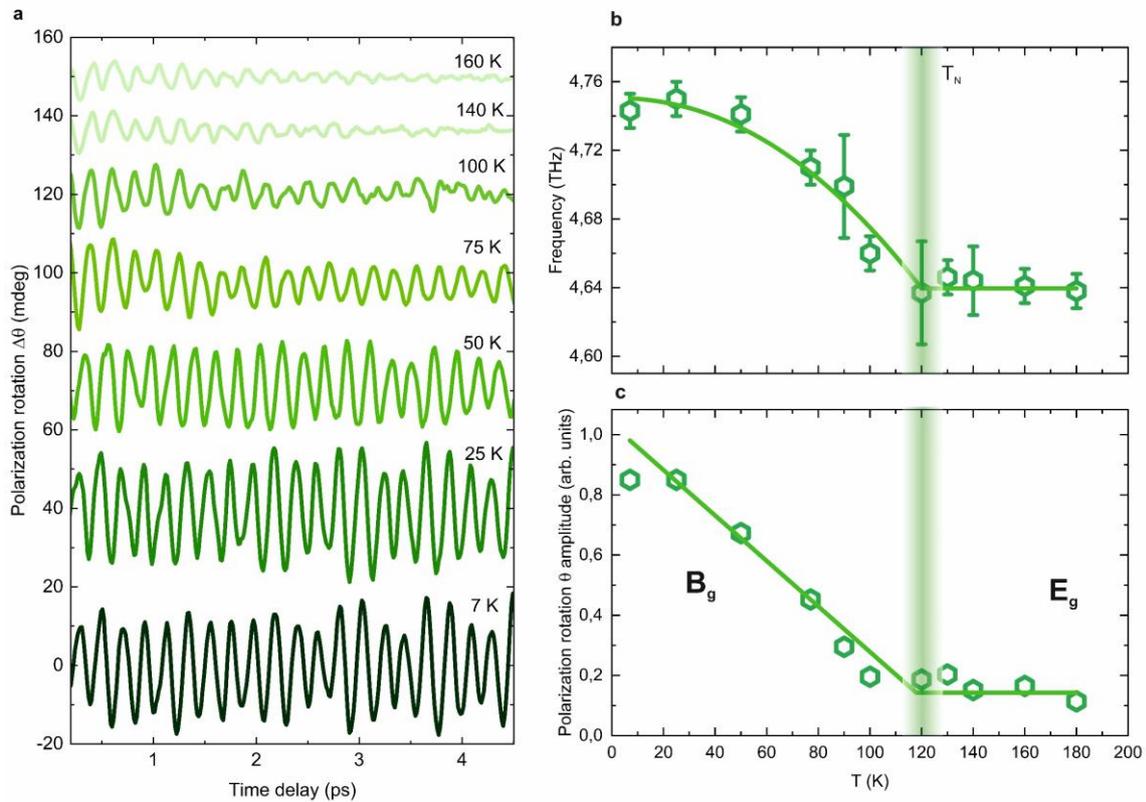

**Figure 4 | High frequency coherent spin-coupled THz phonon mode in CoPS$_3$.** *(a) Time-resolved rotation of the probe polarization showing dynamics of the pump-excited coherent phonon plotted for various temperatures across $T_N$. (b) Frequency f of the phonon mode as a function of temperature T. (c) Amplitude of the phonon mode as a function of temperature T. The solid lines are independent linear fits below and above $T_N$.*

In our experiments, to probe the anisotropy of the optical properties induced by the spin order, we employ the fact that if the polarization of the incoming light does not orient along and/or perpendicularly to the Néel vector (the *a*- and *b*-axis, respectively), the MLD will result in rotation of the polarization plane for light propagating through the sample. In our case, the light, initially polarized at 45 degrees with respect to both the *a*- and *b*-axes, upon propagation through a 4 μm thick sample experiences a net polarization rotation for an angle reaching about 13 degrees (Fig. 1e). We note that, although the value of the angle is large, it is by far less than expected for an MLD as large as 30%. This difference can be explained by the fact that according to the Kramers-Kronig relations[51,52], linear dichroism is always accompanied by linear birefringence. The latter can modify the polarization state of light from linear to elliptical and even circular and thus substantially hampers the measurements of the polarization rotation[53].

To study light-induced ultrafast dynamics of spins and lattice in $CoPS_3$ we carried out an all-optical pump-probe experiment, see Fig. 2a. To excite the dynamics, we employed ultrashort (~50 fs) linearly polarized pulses of light with the photon energy $h\nu$ tunable in a range from 0.7 to 2.5 eV. This energy range covers the vast majority of the orbital *d-d* transitions in $Co^{2+}$ ions[44]. To probe the light-driven spin dynamics, we relied on measuring the polarization rotation $\Delta\theta$ providing, as we have already shown, access to the Néel order. The polarization rotation was also employed to detect the lattice vibrations (phonons), the dynamics of which are intrinsically highly anisotropic and thus contribute to the linear birefringence[17,54]. Figure 2b shows an example of the light-induced dynamics triggered by the pump pulse at a temperature of T=77K when the sample is in the AFM phase. It is seen that, after the excitation, the signal of the transient rotation $\Delta\theta$ suddenly drops, indicating a suppression of the AFM order. Note, that the suppression does not occur instantaneously within the duration of the pump pulse, but rather proceeds on a longer timescale $\tau_s$ of about 1.5 ps (Fig. 2b). The quenching of the AFM order is concomitant with a set of coherent high-frequency phonon oscillations featuring frequencies from 3 to 10 THz and dominated by a phonon mode at 4.74 THz, see Supplementary Materials Fig. S2.

**Spin dynamics**

Figures 3a and 3b show the pump-induced polarization rotation $\Delta\theta$, demonstrating the evolution of the light-induced quenching as a function of the temperature *T*. It is seen that upon approaching $T_N$ the degree of the quenching is gradually growing and peaks right below $T_N$. Above $T_N$ no significant light-induced magnetic dynamics is seen: the small residual signal is likely of non-magnetic origin and caused by pump-induced changes to the electronic part of the dielectric function[55]. The growth of the quenching amplitude is accompanied by the growth of the quenching time $\tau_s$. Upon approaching $T_N$ (Fig. 3c), this time increases from 1.6 ps at T=10 K to a maximum detected time of 5 ps at about T=118 K. We note that qualitatively similar behavior was also observed when the probing was performed in the reflection geometry, see Supplementary Materials S3.

To establish the origin of the AFM quenching, we varied the photon energy of the pump pulse in the range of several *d-d* transitions of $Co^{2+}$ ions. We find that the amplitude of the quenching scales with the absorption and does not depend on the origin of the optical transition. The quenching is defined by the amount of heat deposited into $CoPS_3$ by a laser pulse. The origin of the laser-induced quenching of AFM order is thus similar to that reported earlier for other materials from the *XPS*$_3$ family[17]. Although laser-induced demagnetization is often described using a three-temperature 3*T*-model[1], $CoPS_3$ lacks free electrons and the model is not adequate in this case. Instead, the quenching of the AFM order in $CoPS_3$ can be described by a two-temperature 2T model, where the pump photons increase the potential energy of the excited electrons without making them hot. Upon non-radiative recombination of the excited electrons, the latter promptly (<1 ps) transfer the gained photon energy to the lattice[56]. The subsequent heat exchange via spin-lattice interaction leads to an increase of the effective spin temperature and consequently to a melting of the spin order[57]. This mechanism can explain a substantial increase in the magnitude of the quenching as the temperature approaches $T_N$. Indeed, at higher temperatures the derivative of the Néel order parameter *L* with respect to the temperature increases, and the spins become more susceptible to temperature variations, see Fig. 1d. Remarkably, not only the amplitude of the quenching goes up, but the quenching time

$\tau_s$ also experiences a substantial increase in the vicinity of $T_N$. Such a critical slowing down of the transient spin dynamics induced by light is seen in many AFM compounds, including in recently published results on MnPS$_3$ and FePS$_3$[18,34]. It has been shown[17] that in magnetic insulators the rate of the spin-lattice coupling defining $\tau_s$ scales with the spin-specific heat $C_s$, such that $\tau_s=C_s/g_{sl}$, where $g_{sl}$ is the spin-lattice relaxation rate representing the strength of the spin-lattice interaction. As the heat capacity near the Néel temperature is expected to diverge, assuming a weak temperature dependence of $g_{sl}$ near $T_N$, the characteristic time $\tau_s$ is expected to follow the divergence of $C_s$. Indeed, our experiment shows that $\tau_s \propto |T_N-T|^{-\alpha}$ with a critical exponent $\alpha \sim 0.1$. Remarkably, this value closely matches the theoretical one of $\alpha=0.1$ characterizing the critical scaling of the heat capacity in the 3D Ising model[50,58].

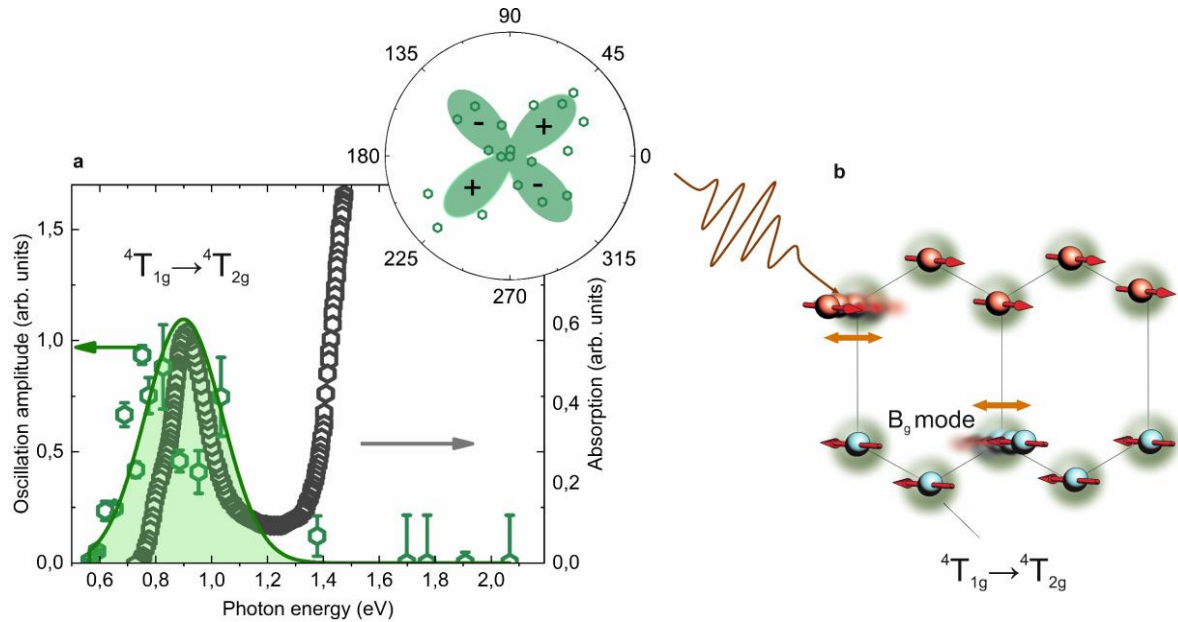

**Figure 5. (a)** *The amplitude of the $B_g$ phonon mode as a function of the pump photon energy (left axis). Optical absorption of CoPS$_3$ in the near IR to the visible range: absorption line at around 0.9 eV corresponds to the $^4T_{1g} \rightarrow {}^4T_{2g}$ orbital transition (right axis). The data is taken from Ref. 44. The samples temperature is 77 K. Inset: Amplitude of the phonon as a function of the pump polarization angle θ with respect to the a-crystal axis. The sign "+" and "-" indicates the relative phase of the oscillations and corresponds to 0° and 180°, respectively.* **(b)** *Schematics of the orbital excitation of the $B_g$ phonon, where blue and red ions are Co$^{2+}$ ions with antiparallel spins.*

We would like to emphasize that the characteristic time for the quenching of the AFM order in CoPS$_3$ ($\tau_s$ =2 ps) is much shorter than the one reported for other XPS$_3$ compounds and MnPS$_3$, in particular ($\tau_s$ =30 ps) [18]. It is well known that the heat capacity $C_s$ scales with the strength of the exchange interaction, which is proportional to the Néel temperature. As CoPS$_3$ and MnPS$_3$ are characterized by similar values of $T_N$, the difference in the quenching time can only be explained by a difference in the magnitude of the spin-lattice relaxation rate $g_{sl}$. Using the available data for the specific heat capacity[15,59], we estimate that $g_{sl}$ to be about $1 \cdot 10^{14}$ Wm$^{-3}$K$^{-1}$ and $5 \cdot 10^{11}$ Wm$^{-3}$K$^{-1}$ for CoPS$_3$ and MnPS$_3$, respectively. Our experiment thus shows that the strength of the coupling can be effectively changed by more than two orders of magnitude by inducing an orbital momentum at the ground state of the magnetic ion. We note that, although the obtained values of the spin-lattice relaxation rates in CoPS$_3$ due to the unquenched momentum of the Co$^{2+}$ ions are high, this is only in comparison with electronically similar materials lacking mobile electrons (dielectrics and semiconductors). In metals, free electrons serve as an additional reservoir of energy that can increase the rate of laser-induced demagnetization by another two orders of magnitude[60].

**Lattice dynamics**

To further understand the nature and excitation mechanism of the light-driven coherent phonon mode that dominates the $\Delta\theta$ signal, we studied its dynamics as a function of both temperature and photon energy. Figure 4a shows the oscillations measured at various $T$. It is seen that the excitation is the most efficient below $T_N$, where both the amplitude and the frequency of the oscillations are strongly dependent on $T$. Using the Fourier transform we retrieved the frequency and amplitude of the oscillations and plotted them as functions of $T$ in Fig. 4b and Fig. 4c, respectively. Upon temperature increase, the phonon frequency $f$ softens until it reaches $T_N$ where it stabilizes at $f_0$=4.64 THz. Fitting the temperature evolution of the frequency shift $\Delta f = f - f_0$ to a critical law similar to the one used for $\alpha_{MLD}$, we find that their critical exponents closely match each other, thus indicating that $\Delta f$ also follows $L^2$. In accordance with Ref.[15], we assign the oscillations to the $B_g$ phonon mode that involves antiphase motions of the magnetic $Co^{2+}$ ions in the direction parallel to the orientation of their spins (Fig. 5 b). The comparison of the temperature behavior of the amplitude of the phonon mode in AFM and PM phases shows that in PM ($T > T_N$) there is no significant temperature dependence but in AFM ($T \leq T_N$) the amplitude rises linearly as temperature decreases. The rise, concomitant with the onset of the AFM order, clearly indicates that the establishment of the Néel order facilitates the excitation of the lattice dynamics.

Figure 5a shows the amplitude of the $B_g$ phonon mode as a function of the pump photon energy $h\nu$. Unlike the light-induced spin quenching the laser-induced lattice dynamics does not correlate with the absorption coefficient but instead is most efficient if the photon energy is in resonance with the $d$-$d$ transition $^4T_{1g} \rightarrow {}^4T_{2g}$ in the $Co^{2+}$ ions at 0.87 eV [44]. We have also found that the efficiency of the excitation is strongly dependent on the incoming pump polarization. The inset in Fig. 5a shows how the amplitude of the phonon oscillations depends on the angle $\gamma$ the pump polarization plane forms with the $a$-axis along which the spins chains are formed. It is seen that the excitation is most efficient if the pump is polarized at 45° with respect to the spins, while no oscillations are excited if the polarization is oriented either along or perpendicularly.

To explain the unusual temperature behavior of the $B_g$ phonon mode and its excitation mechanism we first refer to Ref.[61] which demonstrates that in $FePS_3$, also characterized by unquenched angular momentum, the spin-lattice interaction leads to hybridization of spin and lattice dynamics accompanied by renormalization of their frequencies. The energy of such coupling to the lowest order acquires the following form:

$$\Phi^{sp} = U \cdot Q \cdot lL \quad (2)$$

where $U$ is a phenomenological coupling parameter, $Q$ is the normal coordinate of the $B_g$ mode and $l \ll L$ is a component of the Neel vector along the $b$ – axis, describing its deviation from the ground state $\mathbf{L} = L\hat{\mathbf{a}}$. It is thus clear to enable the spin-lattice coupling in $CoPS_3$ the spins have to be brought out-of-equilibrium. Next, we consider the energy of the light-matter interaction enabling excitation of both the Neel vector $l$ and $B_g$ phonon mode. The energy describing this interaction reads:

$$\Phi^{lm} = (\alpha Q + \beta lL) \cdot E_a E_b, \quad (3)$$

where $E_{a,b}$ are time-dependent electric field components of the pump pulse, and $\alpha,\beta$ are phenomenological parameters, that describe the strength of the light-matter coupling in CoPS$_3$. It is seen that the pump can be coupled not only to the phonon mode but also to the Neel vector. Remarkably, the efficiency of this coupling is proportional to $E_a E_b \propto \sin 2\gamma$, and thus perfectly agrees with the pump polarization dependence of the phonon amplitude, see inset in Fig. 5a. In Supplementary Materials we demonstrate that sufficiently short pump pulse can impulsively excite coherent dynamics of both $Q$ and $l$ in accordance a mechanism known as impulsive stimulated Raman scattering (ISRS)[62,63]. We also show that the inclusion of the spin-lattice coupling hybridizes their dynamics and may explain the enhancement of the phonon amplitude in the AFM phase. Indeed, the hybridization renormalizes the phonon coordinate $\tilde{Q} = Q + \zeta l$. The parameter $\zeta$ characterizes the degree of the hybridization and is proportional to the product $UL$. Therefore, it may naturally explain the enhancement of the phonon excitation in the AFM phase, where $L \neq 0$. Moreover, the hybridization causes a renormalization of the phonon frequency $\Delta f$ in the AFM phase:

$$\Delta f \propto \zeta^2 \propto (UL)^2. \qquad (4)$$

that once again lines up with our experimental findings. We note that although the suggested theory explains experimental findings remarkably well, it still relies on the excitation of the coherent spin dynamics, which is not observed in our experiments. This is rather surprising as excitation and magneto-optical detection of the coherent spin dynamics with femtosecond optical pulses has been reported recently in both MnPS$_3$ and NiPS$_3$ [18,19]. The analogy with the latter is particularly striking as NiPS$_3$ is magnetically isomorphous to the studied here CoPS$_3$. Future probes of the laser-induced spin dynamics in CoPS$_3$ e.g., time-resolved THz spectroscopy, may help to elucidate this issue[64]. In accordance with these theoretical considerations the resonance dependence of the amplitude of the phonon mode on the pump photon energy, shown in Fig. 5a, implies that either both or one of the phenomenological parameters $U, \beta$ are maximized at this photon energy. This implies that, among all possible $d$-$d$ electronic transitions in the range from 0.5 eV to 2.2 eV, the excitation of the B$_g$ mode is most affected by the $^4T_{1g}$ to $^4T_{2g}$ transition, which can be seen as a change of the effective orbital momentum of the magnetic Co$^{2+}$ ions, see Fig. 5b. The corresponding changes to the interionic potential, mediated by the strong spin-lattice coupling, are likely is the mechanical force that moves the ions in the direction parallel to the orientation of their spins.

## Conclusions

To conclude, we have shown that in the vdW AFM CoPS$_3$, characterized by strong spin-lattice coupling, the AFM order can be effectively probed with MLD. Using MLD we detected the ultrafast dynamics of spins and lattice induced by ultrashort pulses of light. We showed that light can suddenly heat the magnetic system, leading to a substantial loss (~1%) of the spin ordering within nearly a single picosecond. Resonantly pumping $d$-$d$ transitions in magnetic Co$^{2+}$ ions, we effectively change the orbital momentum of Co$^{2+}$ ions and show that this excitation, mediated by the spin-lattice coupling, brings Co$^{2+}$ ions in a coherent motion in the direction of the AFM Néel vector. Our experiments not only elucidate the nature of the ultrafast spin-lattice coupling in 2D vdW AFMs but also lay the ground for future ultrafast pump-probe experiments, particularly those aimed at resonant pumping of infrared-active structural phonon modes[6,65].

## Acknowledgments

The authors are grateful to M. Matthiessen, J.R. Hortensius, and A.D. Caviglia for fruitful discussions and to S. Semin, and C. Berkhout for technical support. This work was funded by the Netherlands Organization for Scientific Research (NWO), the European Union Horizon 2020 and innovation program under the European Research Council ERC grant agreement no.856538 (3D-MAGiC), and European Union Horizon 2020 innovation program under the Marie Skłodowska-Curie Grant Agreement No. 861300 (COMRAD), the National Research Fund of Ukraine within project no. 2020.02/026, the Gravitation program of the Dutch Ministry of Education, Culture and Science (OCW) under the research program "Materials for the Quantum Age" (QuMat) registration number 024.005.006, the ERC (Grant No. 1010 78206, ASTRAL), SMV thanks the Generalitat Valenciana for the postdoctoral fellow APOSTD-CIAPOS2021/215

# Supplementary Materials

## Ultrafast laser-induced spin-lattice dynamics in the van der Waals antiferromagnet CoPS$_3$

D. Khusyainov*, T. Gareev, V. Radovskaia, K. Sampathkumar, S. Acharya, M. Šiškins, S. Mañas-Valero, B.A. Ivanov, E. Coronado, Th. Rasing, A.V. Kimel and D. Afanasiev

*Corresponding author. Email: dinar.khusyainov@ru.nl

# 1. Spectral dependence and origin of MLD in CoPS$_3$

Spectral measurements of magnetic linear dichroism (MLD) have been performed on an individual CoPS$_3$ 100 nm thin flake, different from the one used in our time-resolved experiments. The flake was exfoliated from the very same bulk crystal as the one used in the time-resolved experiments. Figure S1a shows a schematic of the experimental setup. The polarization-sensitive absorption measurements have been performed in the transmission geometry. A halogen lamp light source has been employed as a broadband radiation source for spectral measurements. Figure S1b shows the spectral dependence of $\alpha_{\mathrm{MLD}}$ as a function of the sample temperature. A clear difference in the optical absorption of light polarized along the $a$- and $b$-axis is seen in our experiments. Figure S1c demonstrates the spectral dependence of $\alpha_{\mathrm{MLD}}$, as defined in the manuscript by Eq. (1), for various temperatures $T$ below and above $T_\mathrm{N}$. Two pronounced bands in $\alpha_{\mathrm{MLD}}$ are seen at $\lambda_1=530$ nm and $\lambda_2=730$ nm, respectively. Their spectral weight rapidly goes down with temperature and nearly disappears above $T_\mathrm{N}=120$ K, signifying their sensitivity to the spin order.

To infer the physical origin of the bands responsible for the MLD we compare their positions to the electronic transitions defining the optical absorption in the visible range. It is known that in the XPS$_3$ family, similarly to other transition metal complexes, the optical absorption in the visible range is dominated by so-called $d$-$d$ transitions[1–3]. These transitions are electronic transitions that occur between the molecular orbitals of the transition metal ions, like Co$^{2+}$ in CoPS$_3$. Indeed, the position of the first MLD band centered at $\lambda_1=530$ nm closely matches the energy of the orbital $d$-$d$ transition from the $^4T_{1g}$(F) to $^4A_{2g}$(F) states[4]. The position of the second MLD band centered at $\lambda_2=730$ nm matches the energy of another $d$-$d$ transition from $^4T_{1g}$(F) to $^4T_{1g}$(P) states. Therefore, we conclude that the physical origin of the MLD in CoPS$_3$ lies in the $d$-$d$ orbital transitions in the transition metal ion (Co$^{2+}$), widely known to be responsible for the MLD in the visible part of the optical spectrum[5].

We note that the absolute value of the MLD at 800 nm is somewhat smaller than the one reported in our manuscript. This difference can be explained by the linear birefringence that dramatically influences the polarization state of the light as it propagates along the crystal. For instance, it can modify the polarization state of light from linear to elliptical and even circular, thus substantially hampering the polarization-sensitive measurements of the optical absorption[6].

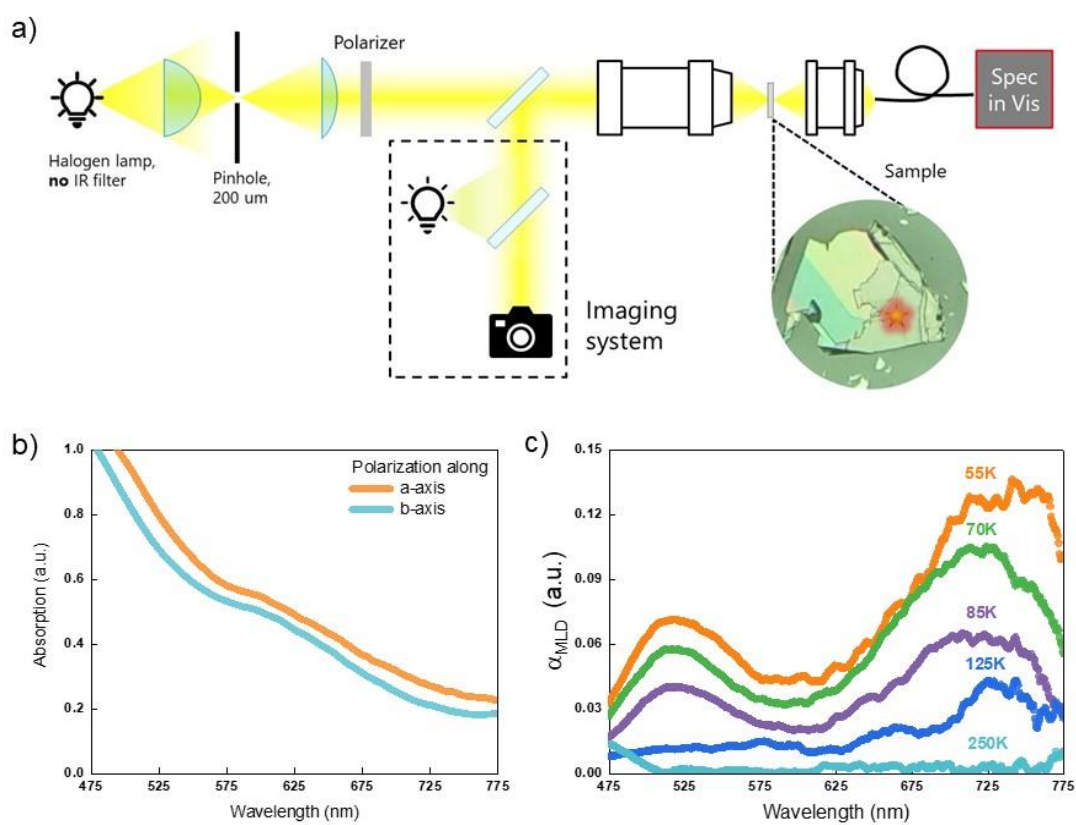

**Figure S1.** Spectral dependence of magnetic linear dichroism (MLD) in $CoPS_3$. **(a)** Schematic diagram of the MLD experiment. The inset shows an optical image of the sample with a probed area marked by a glowing star. **(b)** Spectrally resolved absorption of the sample for the probe light polarized along both the *a-* and *b-* axes. **(c)** MLD spectrum plotted for various temperatures across $T_N$.

## 2. FFT spectrum of the coherent high-frequency oscillations

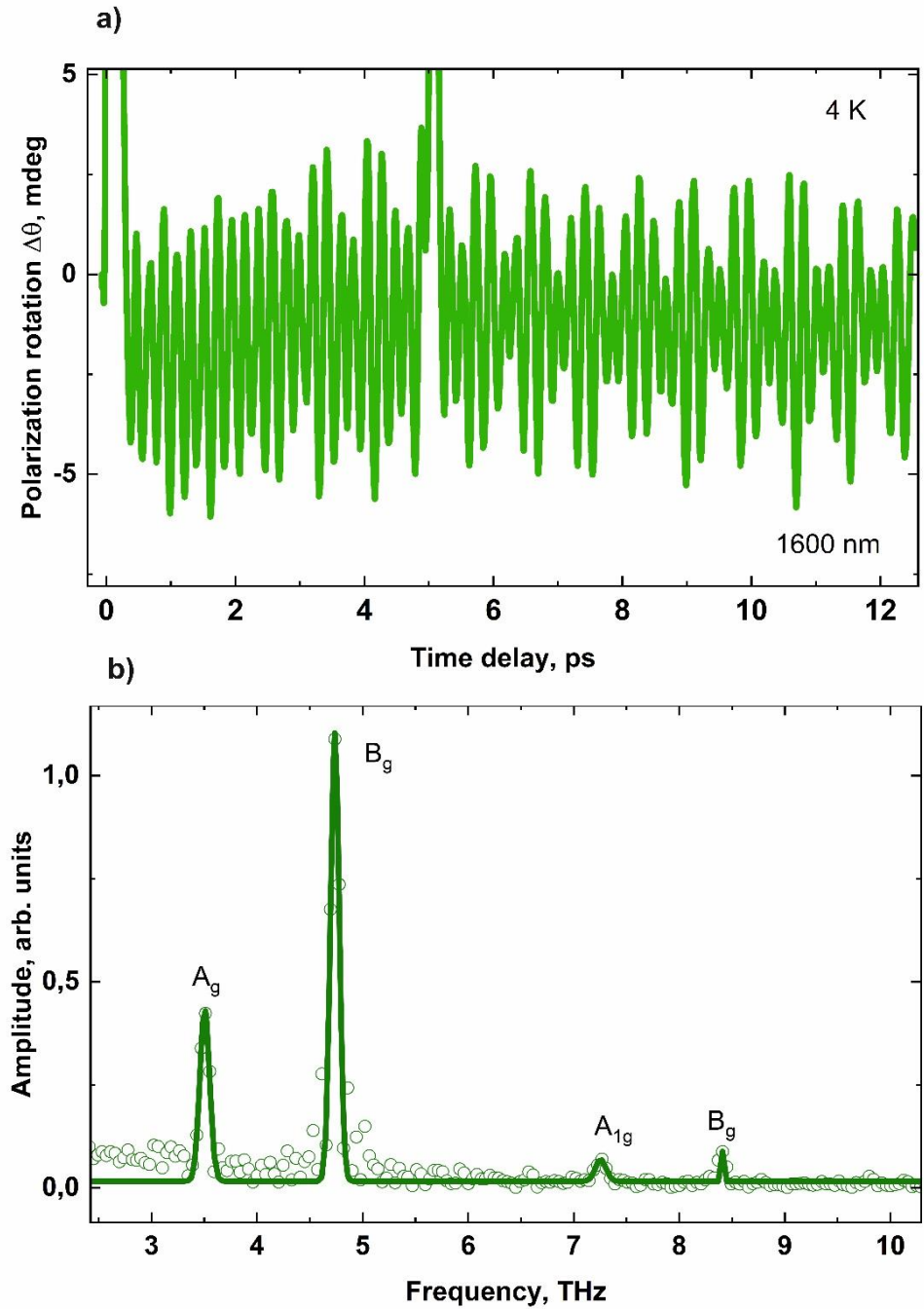

**Figure S2.** (a) Time-resolved coherent ultrafast lattice dynamics in $CoPS_3$ induced by the pump pulse. The exponential background responsible for the incoherent ultrafast melting of the AFM order is removed for clarity. (b) Fast Fourier transform (FFT) of the high-frequency oscillations. Several Raman-active phonon modes of different symmetry are observed and identified according to Ref [7].

## 3. Quenching time and polarization rotation as a function of temperature as measured in the reflection geometry

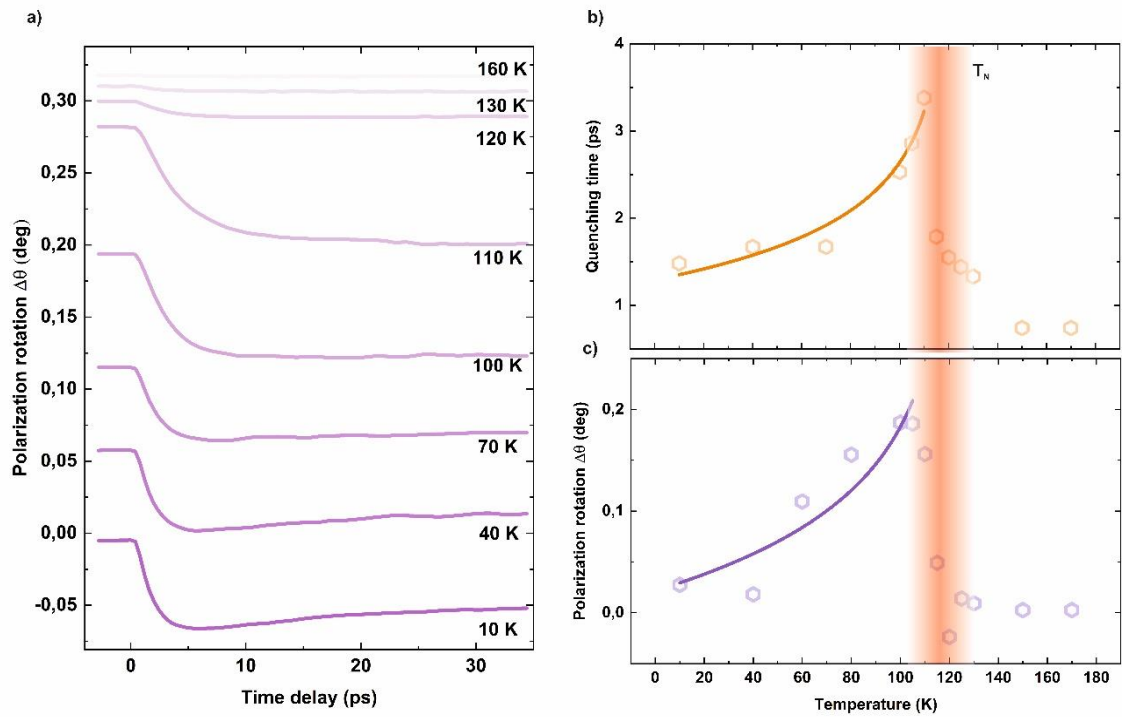

**Figure S3. Quenching of the AFM order as measured in the reflection geometry. (a)** Time-resolved quenching of the antiferromagnetic order as a function of the temperature $T$. No quenching is seen above $T_N$=120 K **(b)** Quenching time $\tau_s$ as a function of $T$. **(c)** Amplitude of the quenching as a function of $T$.

## 4. Spin-lattice coupling constant $G_{sl}$ as a function of temperature

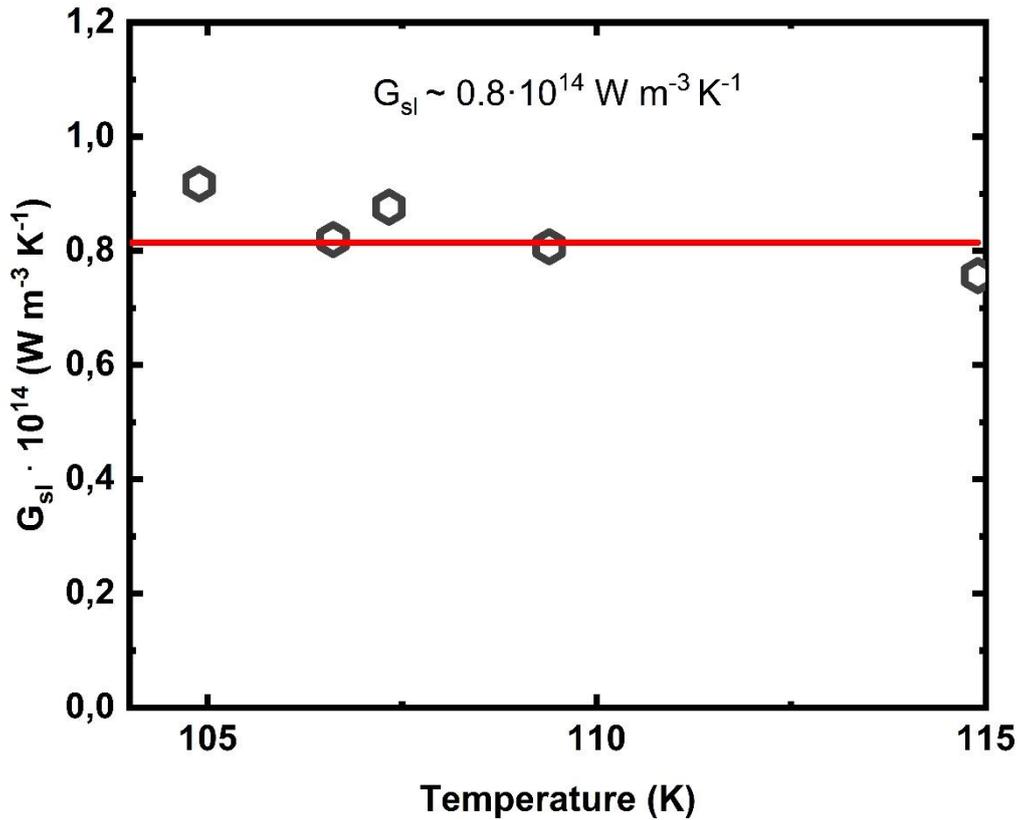

**Figure S4.** Spin-lattice coupling constant $G_{sl}$ as a function of temperature in $CoPS_3$.

## 5. Spin-phonon coupling mechanism.

As $CoPS_3$ is a complex system, which could be explained with 3 exchange interaction integrals, we will use a simple intuitive phenomenological model based on symmetry analysis of the spin-phonon coupling after excitation with short laser pulses. In this way, normal coordinate $Q$ for $Co^{2+}$ ions vibrations along a crystal axis ($B_g$ phonon) and the sigma model for Néel vector **L** oscillation are included in our model in the linear approximation. For the description of Q, it is convenient to use displacements of the ion normalized by atomic spacing. Further, we introduce the coupling term, which could be written as $l$ component obtained after **L** linearization, which is aligned along the b crystal axis multiplied by the normal coordinate $Q$ and the phenomenological constant $U$, which describes spin-phonon coupling (see main text).

By adding a term for optical excitation using the Raman scattering tensor for the $B_g$ phonon for the $C_{2h}$ point group, our system can be described using the following Lagrange function:

$$\mathcal{L} = \frac{M}{2}(\dot{Q}^2 - \omega_0^2 Q^2) + \frac{m}{2}(\dot{l}^2 - \omega_{l0}^2 l^2) - UQlL - (\alpha \cdot Q + \beta \cdot lL)E_a E_b, \qquad (1)$$

where the first two terms describing the motion of the $Q$ and $l$, $M$ is an effective mass of the Q, $m$ is an effective mass of the $l$, $\omega_0$ is the frequency of the phonon mode without coupling, i.e., above Néel temperature, $\omega_{0l}$ is a frequency of the **L** without spin-phonon coupling, when $L$=0, $\alpha, \beta$ and $U$ are phenomenological constants, $E_a, E_b$ are time-dependent electric field components along a and b crystal axis, $L$ is a component of **L** along the *a* crystal axis and $l \ll L$ (see main text). Electric fields component could be described as $E_a E_b = E_0^2(t) \cdot sin2\gamma$, where $E_0(t)$ is the time-dependent amplitude of electric field component, which describes laser pulse and $\gamma$ is an angle between the electric field direction and *a* crystal axis. Using this Lagrangian we obtained differential equations of motion:

$$M(\ddot{Q} + \omega_0^2 Q) + ULl = -\alpha E_0^2(t) \cdot sin2\gamma,$$
$$m(\ddot{l} + \omega_{l0}^2 l) + ULQ = -\beta L E_0^2(t) \cdot sin2\gamma. \qquad (2)$$

Let us start with the analysis of free oscillations of lattice and spins, which happens after the laser pulse action. The general dynamics of the coupled spin-lattice system can be described as a superposition of two normal modes with frequencies $\omega$, which are described by linear homogeneous equations:

$$(\omega^2 - \omega_0^2)Q = ULl/M,$$
$$(\omega^2 - \omega_{l0}^2)l = ULQ/m, \qquad (3)$$

where the frequencies $\omega$ of the normal modes are determined by quadratic over $\omega^2$ equation,

$$(\omega_{ph}^2 - \omega_0^2)(\omega_l^2 - \omega_{l0}^2) - \frac{(UL)^2}{mM} = 0, \qquad (4)$$

with the solution

$$\omega^2 = \frac{1}{2}(\omega_0^2 + \omega_{l0}^2) \pm \frac{1}{2}\sqrt{(\omega_0^2 - \omega_{l0}^2)^2 + 4\frac{(UL)^2}{mM}}, \qquad (5)$$

where the signs plus and minus corresponds to the frequency of lattice-dominated (phonon) mode $\omega_{ph}$ and the spin-dominated (magnon) mode $\omega_l$, respectively. Due to the experimental data, the value of the frequency shift $\omega_{ph} - \omega_0$ is relatively small, and we can use the approximate expression:

$$\omega_{ph} \cong \omega_0 + \frac{(UL)^2}{2\omega_0 mM(\omega_0^2 - \omega_{l0}^2)}, \qquad (6)$$

where the dependence on L coincides with that found in experiment, see Fig 4 b. in the main text. Then using Eq. (4) and the experimental data for $\omega_{ph}$ at low temperatures ($L$=1) and above the Néel temperature ($L$=0) we can estimate the coupling parameter $U$. To do this, we will use

$\omega_{ph} = 4.75$ THz for low temperatures and the value $\omega_0 = 4.64$ THz, and explore $\omega_l = 3.1$ THz according to ref [8] as a magnon gap in CoPS$_3$, effective masses we use as $m = \frac{\hbar}{\omega_{ex}}$ (see e.g. Ref. [9]) with the value of exchange frequency for CoPS$_3$ $\omega_{ex} = 7$ THz,[1] and we estimate $M \sim \frac{\hbar}{\omega_D}$, where $\omega_D$ is the Debye frequency, of the order of $\omega_D \sim 2\omega_0$.[1] Finally, we arrive to the estimate U/$\hbar \approx 0.42$ THz.

Now we can describe $Q$ and $l$ with coupling contribution. Further, we can introduce renormalized normal coordinates $\tilde{Q}$ and $\tilde{l}$ with parameter $\zeta$:

$$\tilde{Q} \cong Q + \zeta l, \qquad \tilde{l} \cong l + \zeta Q, \tag{7}$$

parameter $\zeta$ could be described as:

$$\frac{l}{Q} = \sqrt{\left(\frac{M}{m}\right)\frac{\omega_{ph}^2 - \omega_0^2}{\omega_{ph}^2 - \omega_l^2}} \simeq \sqrt{\frac{2M\omega_{ph}(\omega_{ph} - \omega_0)}{m(\omega_{ph}^2 - \omega_l^2)}}. \tag{8}$$

We could estimate $\zeta = \frac{l}{Q} \approx 0.3$, thus the coupling is strong enough (the standard dimensionless constant of magnon-phonon coupling rarely exceed $10^{-4}$, see e.g.[10]). Considering that $E_0^2(t)$ is strongly localized function so we can replace it with delta function. In this approximation, the dynamics of lattice and spins after the pulse action leads to the free oscillations, which can be described by using non-zero initial condition of the form of $-M\dot{Q}(+0) = \alpha E_0^2(t) \cdot \tau \cdot sin2\gamma$; $-m\dot{l}(+0) = \beta L E_0^2(t) \cdot \tau \cdot sin2\gamma$, where $E_0^2 \cdot \tau = \int_{-\infty}^{\infty} E_0^2(t)dt$. Thus, the presence of spin-lattice hybridization leads to excitation of both modes even if the only one of two phenomenological constants in Eq. (1), $\alpha$ or $\beta$, is non-zero. If the interaction of pump pulse with the media below T$_N$ is dominated by the constant $\beta$, the amplitude of excited phonon should be proportional to the "fraction" of spin oscillations involved to the normal coordinate, i.e., to the parameter $\zeta \sim L$ and the phonon amplitude is proportional to $L^2$. Interaction of the probe pulse with the media contains additional multiplier $L$ and the dependence of the rotation is expected to be proportional to $L^3 \sim (T_N - T)^{0.9}$ that is common to our observation, see Fig. 3 c.